\documentclass[aps,prl,twocolumn,amsmath,amssymb,superscriptaddress]{revtex4}

\usepackage{graphicx}
\usepackage[ansinew]{inputenc}
\usepackage{bm}% bold math
\usepackage{amsfonts}
\usepackage{amssymb}
\usepackage{textcomp}

\newcommand{\ket}[1]{|#1\rangle}

\begin{document}

%\title{Spin-squeezed states for quantum metrology on an atom chip}
%\title{Spin squeezing by chip-based wave function engineering}
%\title{Chip-based wave function control of atomic interactions and entanglement}
%\title{Quantum engineering of spin squeezed states on a microchip}
%\title{Spin squeezing through chip-based control of atomic interactions}
%\title{Controlled generation of entanglement and spin squeezing on a microchip}
%\title{Chip-based generation of entanglement for quantum metrology}
%\title{Atom chip based generation of multi-particle entanglement for quantum metrology}
\title{Atom chip based generation of entanglement for quantum metrology}
%\title{Atom chip based generation of useful entanglement for quantum metrology}
%\title{Chip-based generation of atomic entanglement by wave-function control of interactions}
%\title{Useful entanglement by chip-based wave function control of atomic interactions}
%\title{Useful entanglement on a chip by wave function control of atomic interactions}
%\title{Control of interactions and generation of entanglement on an atom chip}
%\title{Atom chip based control of atomic interactions and entanglement}
%\title{Controlled interactions and generation of multi-particle entanglement on an atom chip}

\author{Max F. Riedel}
\author{Pascal B{\"o}hi}
\affiliation{Fakult{\"a}t f{\"u}r Physik, Ludwig-Maximilians-Universit{\"a}t, Schellingstra{\ss}e~4, 80799~M{\"u}nchen, Germany}
\affiliation{Max-Planck-Institut f{\"u}r Quantenoptik,  Hans-Kopfermann-Stra{\ss}e~1, 85748~Garching, Germany}
\author{Yun Li}
\affiliation{Laboratoire Kastler Brossel, ENS, 24 rue Lhomond,
F-75005~Paris, France} \affiliation{State Key Laboratory of
Precision Spectroscopy, Department of Physics, East China Normal
University, Shanghai~200062, China}
\author{Theodor W. H{\"a}nsch}
\affiliation{Fakult{\"a}t f{\"u}r Physik, Ludwig-Maximilians-Universit{\"a}t, Schellingstra{\ss}e~4, 80799~M{\"u}nchen, Germany}
\affiliation{Max-Planck-Institut f{\"u}r Quantenoptik,  Hans-Kopfermann-Stra{\ss}e~1, 85748~Garching, Germany}
\author{Alice Sinatra}
\email[E-mail: ]{alice.sinatra@lkb.ens.fr}
\affiliation{Laboratoire Kastler Brossel, ENS, 24 rue Lhomond,
F-75005~Paris, France}
\author{Philipp Treutlein}
\email[E-mail: ]{treutlein@lmu.de}
\email[]{philipp.treutlein@unibas.ch}
\affiliation{Fakult{\"a}t f{\"u}r Physik, Ludwig-Maximilians-Universit{\"a}t, Schellingstra{\ss}e~4, 80799~M{\"u}nchen, Germany}
\affiliation{Max-Planck-Institut f{\"u}r Quantenoptik,  Hans-Kopfermann-Stra{\ss}e~1, 85748~Garching, Germany}
\affiliation{Departement Physik, Universit{\"a}t Basel, Klingelbergstra{\ss}e~82, CH-4056~Basel, Switzerland}

\date{\today}

%\begin{abstract}
%\end{abstract}

% insert suggested PACS numbers in braces on next line
%\pacs{}
% insert suggested keywords - APS authors don't need to do this
%\keywords{}

\maketitle

%---version 12, 17.11.2009: control of interactions on atom chip------------------------------------------------------
\textbf{
%Atom chips provide a versatile `quantum laboratory on a microchip' for experiments with ultracold atomic gases in a compact and robust setup \cite{Fortagh07}. 
Atom chips provide a versatile `quantum laboratory on a microchip' for experiments with ultracold atomic gases \cite{Fortagh07}. 
They have been used in experiments on diverse topics such as low-dimensional quantum gases \cite{Hofferberth07}, cavity quantum electrodynamics \cite{Colombe07}, atom-surface interactions \cite{Lin04,Aigner08}, and chip-based atomic clocks \cite{Treutlein04} and interferometers \cite{Wang05,Schumm05}. %,Guenther07,Jo07,Boehi09}. 
%Moreover, a portable atom chip setup was built \cite{Vogel06}, and key components of such systems are now commercially available. %\cite{ColdQuanta}.
A severe limitation of atom chips, however, is that techniques to control atomic interactions and to generate entanglement have not been experimentally available so far. %, mainly because in magnetic traps it is difficult to tune interactions by means of Feshbach resonances.
Such techniques enable chip-based studies of entangled many-body systems and are a key prerequisite for atom chip applications in quantum simulations \cite{Lloyd96}, quantum information processing \cite{diVincenzo00}, and quantum metrology  \cite{Giovannetti04}.
%Such techniques enable chip-based studies of entangled many-body systems and are a key prerequisite for atom chip applications as compact quantum simulators \cite{Lloyd96}, scalable quantum information processors \cite{diVincenzo00}, and portable devices for quantum metrology \cite{Giovannetti04}.
%Such a technique enables studies of entanglement in many-body systems and is a key prerequisite for potential applications of atom chips in quantum simulations, quantum information processing, and quantum metrology.
%Such a technique enables novel studies of entanglement in many-particle systems and is a key prerequisite for applications of atom chips in quantum simulations, quantum information processing, and quantum metrology.
Here we report experiments where we generate multi-particle entanglement on an atom chip by controlling elastic collisional interactions with a state-dependent potential \cite{Boehi09}. 
%This technique also works if Feshbach resonances for tuning of interactions are not available.
We employ this technique to generate spin-squeezed states of a two-component Bose-Einstein condensate \cite{Sorensen01} and show that they are useful for quantum metrology. 
The observed $\bm{-3.7\pm 0.4}$~dB reduction in spin noise combined with the spin coherence imply four-partite entanglement between the condensate atoms \cite{Soerensen01b} and could be used to improve an interferometric measurement %atomic clock
by $\bm{-2.5\pm 0.6}$~dB over the standard quantum limit \cite{Wineland94}. 
%The observed $\bm{-3.7\pm 0.4}$~dB reduction in spin noise combined with the spin coherence imply four-partite entanglement between the atoms in the condensate and could be used to improve the precision of an interferometer %atomic clock
%by $\bm{-2.5\pm 0.6}$~dB beyond the standard quantum limit. 
Our data show good agreement with a dynamical multi-mode simulation \cite{Li09} and allow us to reconstruct the Wigner function \cite{Wigner32} of the spin-squeezed condensate.
The techniques demonstrated here %make entanglement more accessible and 
could be directly applied in chip-based atomic clocks which are currently being set up \cite{Rosenbusch09}. 
}

In the currently emerging field of quantum metrology \cite{Giovannetti04}, multi-particle entangled states such as spin-squeezed states \cite{Esteve08,Appel09,Schleier09} are investigated as a means to improve measurement precision beyond the `standard quantum limit' \cite{Wineland94}. 
This limit arises from the quantum noise inherent in measurements on a finite number of uncorrelated particles and limits today's best atomic clocks \cite{Santarelli99}. 
Atom chips combine exquisite coherent control with a compact and robust setup \cite{Vogel06}, suggesting their use for quantum metrology with portable atomic clocks and interferometers.
Several techniques to create entangled states on atom chips have been proposed \cite{Calarco00,Treutlein06b,Charron06,Zhao07,Li09}, but none has been experimentally realized so far.

The `one-axis twisting' scheme of \cite{Kitagawa93} in principle allows to create a huge amount of entanglement in a two-component Bose-Einstein condensate (BEC) \cite{Sorensen01,Li08,Li09}.
%A scheme that in principle allows to create a huge amount of entanglement in a two-component BEC was proposed in \cite{Kitagawa93,Sorensen01,Li09}. 
In this scheme, an initially separable (non-entangled) state, where each atom is in a superposition of two internal states $\ket{0}$ and $\ket{1}$, dynamically evolves into a spin-squeezed state in which the condensate atoms are entangled. This is due to %internal-state dependent 
atomic interactions that provide a nonlinear term in the Hamiltonian for the BEC internal state. %for the atomic field, 
%realizing the `one-axis twisting' Hamiltonian of \cite{Kitagawa93}.
In the experiments reported here, we realize this scheme on an atom chip. A notable feature is that we control the interactions through the wave function overlap of the two states \cite{Poulsen02,Li09} in a state-dependent microwave potential \cite{Boehi09}.
This is a new and versatile technique for tuning of interactions in a BEC which also works in magnetic traps and for atomic state pairs where no convenient Feshbach resonances exist. We use such a pair, the hyperfine states $\ket{0}\equiv \ket{F=1, m_F=-1}$ and $\ket{1}\equiv \ket{F=2, m_F=1}$ of $^{87}$Rb, which is also employed in chip-based atomic clocks with magnetically trapped atoms \cite{Treutlein04,Rosenbusch09}.

\begin{figure*}[htb]
\includegraphics[width=12cm,clip=]{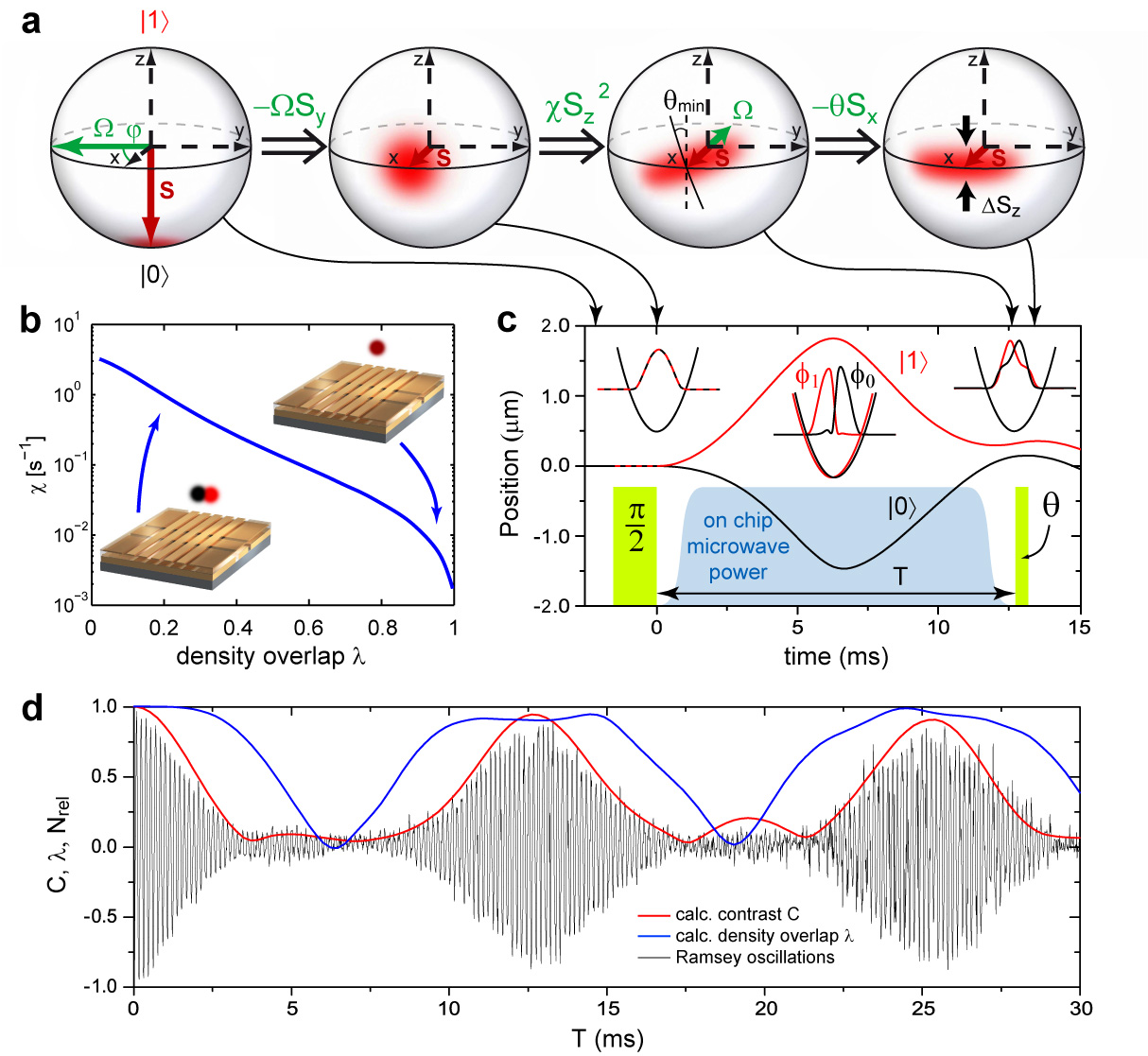}
\caption{\label{fig:overlap}
\textbf{Spin squeezing and entanglement through controlled interactions on an atom chip.}
%\textbf{Realization of the one-axis twisting model for spin squeezing on an atom chip.}
%\textbf{Spin squeezing through controlled nonlinear evolution on an atom chip.}
\textbf{a}, Evolution of the BEC internal state on the Bloch sphere ($\delta=0$ for illustration). Starting with all atoms in $\ket{0}$, a $\pi/2$-pulse prepares a coherent spin state with mean spin $\langle \mathbf{S} \rangle$ along $x$ and isotropic quantum noise in the $yz$-plane (fuzzy red circle). Subsequent nonlinear evolution with %the `one-axis twisting' Hamiltonian 
$\chi S_z^2$ deforms the noise circle into an ellipse, creating a spin-squeezed state with reduced noise at an angle $\theta_\mathrm{min}$. For state tomography, a second pulse rotates the state around $-x$ by a variable angle $\theta$, followed by detection of $S_z$.
%To measure the spin quadrature $S_\theta=(\cos \theta) \, S_z - (\sin \theta) \, S_y$, a second pulse rotates the state around $-x$ by a variable angle $\theta$, followed by detection of $S_z$.
\textbf{b}, Control of the nonlinearity $\chi$ on the atom chip. $\chi$ depends on the difference of intra- and inter-state atomic interactions. Its dependence on the normalized density overlap $\lambda$ of the two BEC components is shown, calculated from stationary mode functions in potentials of increasing separation. 
%with a separation between $0~\mu$m ($\lambda=1$) and $xxx~\mu$m ($\lambda=0$).
\textbf{c}, Experimental sequence and motion of the two BEC components corresponding to \textbf{a}. In between the pulses for internal-state manipulation (green), a state-dependent microwave potential is turned on (blue; pulse durations and microwave ramp times exaggerated for clarity). It dynamically splits and recombines the two BEC components, so that $\chi>0$ during the time $T$. The simulated center-of-mass motion of the two states $\ket{0}$ (black) and $\ket{1}$ (red) is shown as a function of time. A slightly asymmetric splitting of the potentials results in an asymmetric oscillation. Insets: corresponding BEC mode functions $\phi_0$ and $\phi_1$ along the splitting direction %in their respective potentials 
at the beginning, in the middle, and at the end of the sequence.
\textbf{d}, Measured Ramsey fringes in the normalized population difference $N_{\mathrm{rel}}$ (here, $\theta=\pi/2$). 
Between the pulses, $\delta$ is nonzero (see Supplementary Information), resulting in spin precession.  
The splitting and recombination of the BEC modulates the fringe contrast.
Here, $T$ is scanned beyond the duration of the sequence in \textbf{a}+\textbf{c}, so that up to $\sim 2.5$ oscillations are induced.
The simulated contrast $C$ (red) and density overlap $\lambda$ (blue) are shown for comparison.
}
\end{figure*}

A two-mode model \cite{Li09} provides a starting point to understand how squeezing is created in our experiment. 
The internal state of a BEC of $N$ two-level atoms can be described by a collective spin $\mathbf{S}=\sum_{i=1}^N \mathbf{s}_i$, the sum of the individual spins 1/2 of each atom (see Fig.~\ref{fig:overlap}a). % of components $S_x=(a^\dagger b+b^\dagger a)/2$, $S_y=(a^\dagger b-b^\dagger a)/(2i)$, and $S_z=(a^\dagger a-b^\dagger b)/2$. 
Its component $S_z=(N_1-N_0)/2$ is half the atom number difference between the states and thus directly measurable.
%; a more accurate dynamical multi-mode model is described in  \cite{Li09}. 
%A more accurate dynamical multi-mode mode accounting for both spatial and spin dynamics is developed in \cite{Li09}.
%The internal state of a BEC of $N$ two-level atoms can be described by a collective spin (sum of the individual spins 1/2 of each atom) of components $S_x=(a^\dagger b+b^\dagger a)/2$, $S_y=(a^\dagger b-b^\dagger a)/(2i)$, and $S_z=(a^\dagger a-b^\dagger b)/2=(N_1-N_0)/2$. Here, $a^\dagger$ ($b^\dagger$) and $a$ ($b$) are creation and annihilation operators and $N_1$ ($N_0$) is the number of atoms in state $\ket{1}$ ($\ket{0}$).
A $\pi/2$-pulse applied to a BEC in $\ket{0}^{\otimes N}$ prepares it in a coherent spin state $(\ket{0}+\ket{1})^{\otimes N}/2^{N/2}$ with mean spin $\langle S_x \rangle = N/2$ and $\langle S_y \rangle = \langle S_z \rangle =0$. This is a product state in which the atoms are uncorrelated and the quantum noise is evenly distributed among the spin components orthogonal to the mean spin, $\Delta S_y^2 = \Delta S_z^2 = N/4$, satisfying the Heisenberg uncertainty relation $\Delta S_y \Delta S_z = |\langle S_x \rangle|/2$.
This noise gives rise to the standard quantum limit if the state is used in a Ramsey interferometer such as an atomic clock \cite{Santarelli99}.
 
Quantum correlations between the atoms can reduce the variance of one spin quadrature in the $yz$ plane at the cost of increasing the variance of the orthogonal one, resulting in a spin-squeezed state \cite{Kitagawa93}.
To quantify its usefulness for metrology, one introduces the squeezing parameter \cite{Wineland94}
$\xi^2=N \Delta S_{\theta,\mbox{\small min}}^2/\langle S_x \rangle^2,$  
where $\Delta S_{\theta,\mbox{\small min}}^2$ is the minimal variance of the spin in the  $yz$ plane (see Fig.~\ref{fig:overlap}a).
The normalization by $\langle S_x \rangle^2$ takes into account that improving interferometric sensitivity requires not only reducing noise but also maintaining high interferometer contrast $C= 2 |\langle S_x \rangle|/N$.
A state with $\xi^2 < 1$ allows one to overcome the standard quantum limit in a Ramsey interferometer by a factor $\xi$ with respect to the use of an uncorrelated ensemble of atoms \cite{Wineland94}.
%We thus want $\xi^2$ to be as small as possible.
Furthermore, $\xi^2$ is an entanglement witness, with $\xi^2<1$ indicating at least bipartite entanglement between the condensate atoms \cite{Sorensen01}.

We produce spin-squeezed states by means of time evolution through the `one-axis twisting' Hamiltonian \cite{Kitagawa93}:
\begin{equation}
H/\hbar =   \delta S_z + \Omega S_\varphi + \chi S_z^2, \label{eq:H}
\end{equation}
which describes our BEC in good approximation \cite{Li09}.
The first term in (\ref{eq:H}) describes spin precession around $z$ at the detuning $\delta$. % including BEC mean field corrections \cite{Li09}. 
The second term describes spin rotations around an axis $S_\varphi=(\cos \varphi) \, S_x - (\sin \varphi) \, S_y$ due to a coupling of $\ket{0}$ and $\ket{1}$ with Rabi frequency $\Omega$ and phase $\varphi$.
%The third, nonlinear term of strength $\chi$ arises due to elastic collisional interactions in the BEC and is responsible for spin squeezing. 
The third, nonlinear term of strength $\chi$ arises due to elastic collisional interactions in the BEC. It `twists' the state on the Bloch sphere (see Fig.~\ref{fig:overlap}a), resulting in spin squeezing and entanglement.

An essential feature of our experiment is the control of this nonlinearity. %It should be active during a well chosen best squeezing time to avoid `oversqueezing', which results in a decrease of $\langle S_x \rangle$ that makes $\xi^2$ finally increase again \cite{Kitagawa93}.
Its coefficient
\begin{equation}
\chi=\frac{1}{2\hbar}\left(\partial_{N_0}\mu_0+\partial_{N_1}\mu_1
-\partial_{N_1}\mu_0-\partial_{N_0}\mu_1\right)_{\langle N_0\rangle,\langle N_1\rangle}
\label{eq:chi}
\end{equation}
depends on derivatives of the chemical potentials
\begin{equation}
\mu_j=\langle\phi_j|h_j|\phi_j\rangle+\sum_{k=0,1} g_{jk}N_k\int dr^3 |\phi_j|^2|\phi_k|^2
\end{equation}
of the two BEC components evaluated at the mean atom numbers $\langle N_0\rangle=\langle N_1\rangle=N/2$ after the $\pi/2$-pulse. Here, $h_j$ is the single-particle Hamiltonian including kinetic energy and the trapping potential, and $\phi_j(\mathbf{r})$ is the spatial mode function of state $\ket{j}$. The interaction strength $g_{jk}=4\pi\hbar^2 a_{jk}/m$ between atoms in $\ket{j}$ and $\ket{k}$ depends on the corresponding s-wave scattering length $a_{jk}$.
For our states, the three scattering lengths are close, $a_{00}:a_{01}:a_{11}=100.4:97.7:95.0$. % \cite{Mertes07}.
If the two BEC modes overlap spatially, $\phi_1 = \phi_0$, the crossed terms in (\ref{eq:chi}) with the minus sign compensate the direct terms with the plus sign. Thus, by default, $\chi \approx 0$.
%It has been suggested that a Feshbach resonance could be used to tune the interaction strengths. However, for our magnetically trappable state pair, no such resonance is available.
In order to increase $\chi$, we control the overlap of $\phi_0$ and $\phi_1$ with a state-dependent trapping potential. 
By spatially separating the two modes, the crossed terms $\partial_{N_1}\mu_0$ and $\partial_{N_0}\mu_1$ are set to zero and thus $\chi>0$. In Fig.~\ref{fig:overlap}b, $\chi$ is shown as a function of the normalized density overlap $\lambda=\int dr^3 |\phi_0|^2|\phi_1|^2 /\sqrt{ \int dr^3 |\phi_0|^4\int dr^3 |\phi_1|^4}$, calculated from stationary mode functions in traps of increasing separation for our experimental parameters (see below). 
%Trap parameters and $N$ are the same as used in the experiment (see below). %Note that $\chi$ can be tuned over three orders of magnitude.
%By spatially splitting and recombining the condensate modes, we can thus turn the nonlinearity on an off. 

%An overview of our experimental setup is given in the Supplementary information, for a more detailed description see \cite{Boehi09}.
Our experimental setup is described in detail in \cite{Boehi09} (see also Supplementary Information).
In short, we use an atom chip to prepare pure BECs of $N = 1250\pm 45$ atoms in state $\ket{0}$ in a harmonic magnetic trap with longitudinal (axial) trap frequency $f_\mathrm{long}=109~$Hz ($f_\mathrm{ax}=500~$Hz). We couple $\ket{0}$ and $\ket{1}$ with $\Omega/2\pi=2.1$~kHz using microwave+rf radiation. The trap minima for $\ket{0}$ and $\ket{1}$ can be shifted with respect to each other along the longitudinal direction using a chip-based state-dependent microwave potential. 
%We detect both $N_0$ and $N_1$ in a single shot of the experiment using state-selective absorption imaging.
For detection, we use state-selective absorption imaging with a carefully calibrated imaging system (see Supplementary Information). It allows us to detect both $N_0$ and $N_1$ with good accuracy in a single experimental run.

Our experimental sequence for squeezing (Fig.~\ref{fig:overlap}a+c) starts with a resonant $\pi/2$-pulse of duration $120~\mu$s %and $\varphi=\pi/2$ 
to prepare the coherent spin state. During the pulse,  $\Omega \gg \chi N$ so that the nonlinearity can be neglected.
After the pulse, we squeeze the state by turning on $\chi$ for a well-defined time by spatially splitting and recombining the two components of the BEC in the following way (see Fig.~\ref{fig:overlap}c). The microwave potential is turned on within $50~\mu$s resulting in an abrupt separation of the trap minima for $\ket{0}$ and $\ket{1}$ by $s = 0.52~\mu$m. The two components of the BEC start to perform one oscillation in their respective potentials. 
During the oscillation, which is strongly influenced by mean-field effects, the mode functions $\phi_0$ and $\phi_1$ almost completely separate so that $\chi = 1.5~\mathrm{s}^{-1}$ at maximum separation. 
After a time $T = 12.7~$ms the states overlap again, the microwave potential is switched off within $50~\mu$s, and the squeezing dynamics, as well as the relative atomic motion, stops. 
This value of $T$ nearly coincides with the `best squeezing time' expected from the two-mode model \cite{Li09}.
We analyze the produced state by performing state tomography. 
With the mean spin along $x$, we measure the transverse spin components $S_\theta=(\cos \theta) \, S_z - (\sin \theta) \, S_y$ along any angle $\theta$ by rotating the state vector in the $yz$-plane by that angle prior to detection of $S_z=(N_1-N_0)/2$. This is done by applying a second pulse for a duration $\tau_\theta= \theta/\Omega$ and with a phase $\varphi = \pi$ ($\varphi=0$) for turning clockwise (counterclockwise). 

%To analyze the produced state, we apply a second pulse ($\Omega \gg \chi N$) of variable area $\theta$ and phase $\varphi$ followed by detection of $S_z$. In this way, we perform tomography of the spin-squeezed state, which allows us to reconstruct its Wigner function and to determine $\xi^2$.
%In step three, we perform tomography of the spin-squeezed state in order to reconstruct its Wigner function and to determine $\xi^2$.
%By detecting $N_0$ and $N_1$ we can directly determine $S_z=(N_1-N_0)/2$. To measure spin components $S_{\theta}$ along any angle $\theta$ we turn the state vector in the $yz$-plane by that angle prior to detection. This is done by applying a second pulse for a duration $\tau_\theta= \theta/\Omega$ and with a phase $\varphi = 0$ ($\varphi=\pi$) for turning clockwise (counterclockwise) ???. 
%%The phase $\varphi$ is adjusted by precise timing of the second pulse,.... mention detuning?

\begin{figure}[htb]
    \centering
        \includegraphics[width=1\columnwidth]{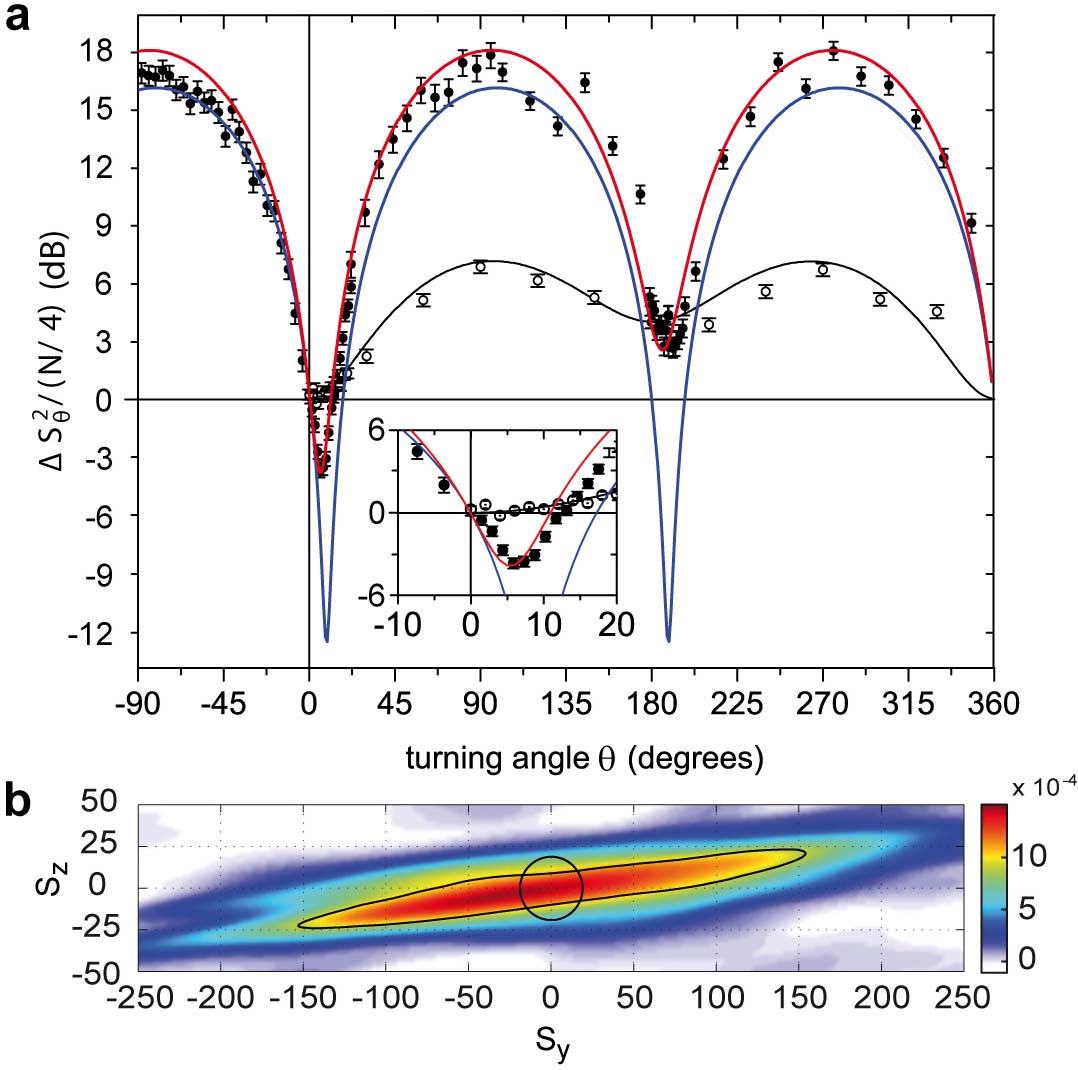}
    \caption{\textbf{Spin noise tomography and reconstructed Wigner function of the spin-squeezed BEC.}
\textbf{a}, Observed spin noise for the spin-squeezed state (solid circles) and for a coherent spin state (reference measurement, open circles). The normalized variance $\Delta_n S_\theta^2=4\, \Delta S_\theta^2/\langle N\rangle$ is shown as a function of the turning angle $\theta$ in the $yz$-plane, with statistical error bars. For this graph, we remove photon shot noise due to the imaging process as described in the Supplementary Information. In the squeezed state, a spin-noise reduction of $-3.7\pm 0.4~$dB is observed for $\theta_\mathrm{min}=6^\circ$, corresponding to $\xi^2=-2.5\pm 0.6$~dB of metrologically useful squeezing for our Ramsey contrast of $C=(88\pm 3)\,\%$. Solid lines are results from our dynamical simulation. Blue: squeezed state with losses but without technical noise; red: squeezed state with losses and technical noise; black: reference measurement with losses and technical noise.
\textbf{b}, Wigner function of the spin-squeezed BEC reconstructed from our measurements. The black contour line indicates where the Wigner function has fallen to $1/\sqrt{e}$ of its maximum. Squeezed and `anti-squeezed' quadratures are clearly visible. For comparison, the circular $1/\sqrt{e}$ contour of an ideal coherent spin state is shown. The area of the contour line is larger than the area of the circle indicating that the squeezed state is not a minimum uncertainty state anymore. %This is due to the technical phase noise.
}
    \label{fig:StdDev}
\end{figure}

Figure \ref{fig:StdDev}a shows the %spin noise 
noise in $S_\theta$ 
obtained from a large number of such measurements as a function of $\theta$. Data for a squeezed state is shown in comparison with data for a coherent spin state where the traps were not separated during the sequence (reference measurement). 
We plot the normalized variance $\Delta_n S_\theta^2=4\, \Delta S_\theta^2/\langle N\rangle$, % as a function of $\theta$. 
so that $\Delta_n S_\theta^2=0$~dB corresponds to the standard quantum limit. 
In the squeezed state, the spin noise $\Delta_n S_\theta^2$ falls significantly below the standard quantum limit, % for small $\theta$, 
reaching a minimum of $\Delta S_\theta^2=-3.7\pm 0.4~$dB at $\theta_{\mathrm{min}}=6^\circ$. 
The corresponding interference contrast is $C=(88\pm 3)\,\%$. This results in a squeezing parameter of $\xi^2=-2.5\pm 0.6$~dB, proving that the state is useful for quantum metrology and that the condensate atoms are entangled. 
The reference measurement, by contrast, stays above the standard quantum limit for all values of $\theta$. 
%Technical phase noise leads to an increase in $\Delta_n S_\theta^2$ around $\theta=90^\circ$ and $\theta=270^\circ$ (see below).

Entanglement in the BEC has been defined as the non-separability of the $N$-particle density matrix \cite{Sorensen01,Soerensen01b}. 
An intriguing question concerns the depth of entanglement: How large must the clusters of entangled atoms be at least in order to produce the observed squeezing?
%Using the procedure described in \cite{Soerensen01b}, we find that the spin noise reduction and mean spin length observed in our experiment proves that the condensate atoms are entangled in clusters of at least four particles.
In \cite{Soerensen01b}, a procedure to determine the depth of entanglement from the measured spin noise reduction and mean spin length is described.
Our data falls below the spin-$3/2$-curve in Fig.~1 of \cite{Soerensen01b}, which is an experimental proof that the condensate atoms are entangled in clusters of at least $4\pm 1$ particles. 
%which is an experimental proof of four-particle entanglement.

%The reference measurement starts out at $\Delta_n S_\theta^2=1$ for small $\theta$ and increases to $\Delta_n S_\theta^2\approx $ at $\theta \approx 90^\circ$ and again at $\theta \approx 270^\circ$ due to technical phase noise. For the squeezed state, on the other hand,  $\Delta_n S_\theta^2$ falls significantly below the standard quantum limit for small $\theta$ and shows an 8-fold increase around $\theta \approx 90^\circ$ and $\theta \approx 270^\circ$. A maximum suppression of $\Delta S_\theta^2$ of $-3.7\pm 0.4~$dB is achieved at an angle of $\theta_{\mathrm{min}}=6^\circ$.

During the splitting, internal and motional states of the atoms are entangled. 
%This can lead to a decrease of the fidelity of the squeezed state if the wave function overlap after recombination is less than unity.
This can lead to a decrease of $C$ and thus an increase of $\xi^2$ if the recombination is not perfect. % if the wave function overlap after recombination is less than unity.
To find the time of maximum contrast, we perform a Ramsey measurement where the second pulse area is $\theta=\pi/2$ and $T$ is scanned.
Figure \ref{fig:overlap}d shows the resulting Ramsey fringes in the normalized population difference $N_\mathrm{rel}=(N_1-N_0)/N$ as a function of $T$. 
%The Ramsey fringe contrast $C$ directly measures the length of the mean spin
%\begin{equation}
%| \langle S \rangle |/(N/2) =C=(\max(N_{\mathrm{rel}})-\min(N_{\mathrm{rel}})/2,
%\end{equation}
%which is sensitive to the wave function overlap at the time of the second pulse (how?). 
We observe a high contrast of $C=(88 \pm 3)\,\%$ at $T=12.7$~ms, indicating large spatial overlap and nearly vanishing relative motion of the two states. 
In the squeezing sequence, we turn off the microwave potential at this time, preserving the large overlap for subsequent measurements. 
%and consequently choose this time for the squeezing measurements. 
The contrast could be further increased using optimal control techniques \cite{Treutlein06b}. % instead of abrupt switching of the potentials.
In comparison with the data, we show $C$ and $\lambda$ as obtained from a simulation for our experimental parameters. 
For an accurate description of our system, accounting for both the spatial and the spin dynamics, we use the dynamical multi-mode theory developed in \cite{Li09}. It neglects initial thermal excitations and reduces to the simple two-mode model described above for the case of stationary condensates. 
The only adjustable parameter in the simulation is the splitting distance $s$, which is not resolved by our imaging system. The resulting value $s = 0.52~\mu$m is consistent with a simulation of the potential. We observe very good agreement with the measurement, indicating that the simulation correctly accounts for the motion of the BEC in the trap. In Fig~\ref{fig:overlap}c, the simulated dynamics of the mode functions $\phi_i$ is shown.

%At the time $T_{R,0}$, $N_{\mathrm{rel}}=0$.
%%(possible values for $T_{R,0}$ are marked with red circles in the inset of figure \ref{fig:overlap}).
%If $T_{R,0}$ corresponds to a rising (falling) slope of the Ramsey fringe the Rabi vector is aligned anti-parallel (parallel) with the state which is therefore turned (anti-)clockwise (\textit{check this!}).
%For an accurate description of the system, accounting for both the spatial and the spin dynamics, we use the dynamical multi-mode model developed in \cite{Li09}. It is valid for $T\simeq 0$ and reduces to the simple two-mode model described earlier in the particular case of stationary condensates. On the other hand we use the stationary Hamiltonian (\ref{eq:H}) in a Master Equation to evaluate the effect of particle losses (1,2,3-body) on the squeezing in our experiment, using the formalism developed in \cite{Li08, Li09}.

The spin noise reduction obtained from this simulation is shown in Fig.~\ref{fig:StdDev}a along with the data. We add the effect of particle loss (1,2,3-body) as determined from the two-mode Hamiltonian (\ref{eq:H}) in a Master Equation approach \cite{Li08, Li09}, as well as several technical noise sources.
The blue line shows the expected squeezing taking into account atom loss but no technical noise. The maximal reduction in variance is $-12.8$~dB, significantly larger than observed. %We attribute this to several technical noise sources, which are included in the red solid line. % leading to a maximally achievable squeezing of $\xi^2=-xx\,$dB taking into account the reduced contrast of xx\,\% in the simulation (compare figure \ref{fig:overlap}).
The red line, which describes our data well, additionally includes the fluctuations of $N$, fluctuations of the pulse power of $0.5\,\%$ r.m.s.,
%a systematic detuning of the two photon drive used for state preparation by $300\,$Hz,
a fluctuating detuning of $2\pi \times 40~$Hz r.m.s.\ during the pulses, and phase noise of $\Delta \varphi=8^\circ$ r.m.s.\  (see Supplementary Information). All fluctuations are consistent with independent measurements.
The fluctuating detuning is due to fluctuating microwave level shifts during the pulses. % caused by fluctuations of the microwave power in the two photon drive.
It is the cause for $\Delta_n S_\theta^2 > 0$~dB at $\theta=180^\circ$.
The phase noise is the main reason why the maximum achieved squeezing is smaller than the theoretically predicted value. 
It is consistent with technical fluctuations of the magnetic trap position in the inhomogeneous microwave near-field. Consequently, the phase noise in the reference sequence is smaller ($\Delta \varphi = 3^\circ$, black line).

%The measured histograms of $S_\theta^2$ for different angles $\theta$ represent a tomography of the squeezed state.
%which is conceptually similar to optical homodyne tomography of squeezed light \cite{Vogel89}.
%Our allows us to reconstruct the Wigner function of the squeezed condensate.
The measured histograms of $S_\theta$ for different angles $\theta$ are tomographic data that allow us to reconstruct the Wigner function $W(S_y,S_z)$ of the squeezed BEC \cite{Wigner32} using the inverse Radon transform (see Supplementary Information).
Figure \ref{fig:StdDev}b shows the reconstruction. % Wigner function of the squeezed state.
The two contour lines indicate where the Wigner functions of our squeezed state and of an ideal coherent spin state (with the same $N$ and
%as it would be detected with our imaging system
with added imaging noise) have fallen to $1/\sqrt{e}$ of their maximum. 
%For the non-squeezed state the contour line is a circle with radius 19.1 which is slightly larger than $\sqrt{N/4}$ due to imaging noise.
%with radius $(\langle N\rangle /4+(\Delta N^{\mathrm{psn}})^2)^{1/2}=19.1$ including the imaging noise of $\Delta N^{\mathrm{psn}}=7.1$ atoms which is present in the squeezing measurement.
The squeezing along the direction $\theta_\mathrm{min}$ as well as the `anti-squeezing' in the perpendicular direction can be clearly seen. 
%The major semi axis encloses the angle $\theta_\mathrm{min}$ with the abscissa, which corresponds to the equator on the Bloch-sphere.
%Quantum state tomography is of interest as it gives access to measures of entanglement 
%%than $\xi^2$  which are not accessible by a simple measurements at one angle $\theta$,
%such as the quantum Fisher information \cite{Frieden95, Pezze09}, which characterizes a more general class of states (including states with $\xi^2>1$) which can be used to overcome the standard quantum limit \cite{Pezze09}.

%Note that the state is not centered around the origin indicating that the mean spin was not perfectly aligned with the Rabi vector during measurement. Possible reasons for a deviation in the $z$-direction include losses and inaccuracy of the first $\frac{\pi}{2}$-pulse area, a deviation in the $y$-direction occurs if $T_R,0$ was not determined exactly (here, already sub-percent deviations lead to a recognizable deviation).

In conclusion, using a novel method to control interactions with a state-dependent potential, we have demonstrated for the first time spin squeezing and multi-particle entanglement on an atom chip. %Our method can in principle produce squeezing with $\xi^2<-12$~dB but is currently limited by technical phase noise. 
We envisage the implementation of this technique in portable atomic clocks and interferometers operating beyond the standard quantum limit.
Furthermore, it is a valuable tool for experiments on many-body quantum physics and could enable quantum information processing on atom chips \cite{Treutlein06b}.

%Furthermore, our method to produce entanglement can be used to measure the interaction-induced phase shift of two colliding BECs exploring the possibility of using BECs for quantum information processing and, together with on-chip single atom detectivity \cite{Steinmetz06}, providing the basis for a quantum phase gate on an atom chip.

%Outlook: our sequence realizes a nonlinear beam splitter. It is the first step in a Ramsey interferometer with squeezed states. what kind of sequence would have to be applied to make use of the squeezing in an atomic clock? Rotate such that small axis of ellipse is along equator of Bloch sphere, then wait to accumulate phase shift to be measured, then another pi/2-pulse. Maybe reference Oberthaler-paper here? Long lifetimes of squeezing after creation by relaxing the trap.

%Future experiments in this direction may shine new light on the still puzzling nature of multi-particle entanglement in systems with a large number of particles.

%- how does our mechanism to produce squeezing scale with atom number?

%- How long can the squeezing be maintained in the presence of losses? (measurement or theoretical estimate?)

%- Can we already make a statement about whether our system can be really useful and competitive for precision measurement (requiring large atom numbers)? Or is it at the moment mostly interesting for fundamental reasons?

The group around M. Oberthaler has independently and simultaneously realized spin squeezing through Fesh-bach control of interactions in an optical trap.

%The group around P. Treutlein has independently and simultaneously realized spin squeezing on an atom chip through controlled interactions with a state-dependent potential.

\section*{Acknowledgements}
It is a pleasure to thank K. M{\o}lmer, J. Reichel, A. Smerzi, and A. S{\o}rensen for helpful discussions and J. Halimeh for careful reading of the manuscript.
This work was supported by the Nanosystems Initiative Munich. T.W.H.\ gratefully acknowledges support by the Max-Planck-Foundation.

\section*{Author information}
Correspondence and requests for materials should be addressed to P.T. (experiment) and A.S. (simulation).

\section*{Author contributions}
A.S. and P.T. jointly conceived the study. 
M.F.R., P.B., and P.T. carried out the experiment and analyzed the data. 
Y.L. and A.S. carried out the simulations. 
All authors discussed the results and contributed to the manuscript.

%\section*{Competing financial interests}
%The authors declare that they have no competing financial interests.

\newpage
\onecolumngrid
\section[Supplementary Information]{Atom chip based generation of entanglement for quantum metrology: Supplementary Information}
\twocolumngrid
\subsection{Experimental setup}

The core of our experiment is an atom chip with integrated microwave guiding structures. It allows us to generate state-dependent microwave near-field potentials in addition to static magnetic traps. The chip, the preparation of Bose-Einstein condensates (BECs), and the use of microwave near-field potentials for state-dependent coherent manipulation of ultracold atoms is described in detail in \cite{Boehi09a}. We briefly summarize it in the following, highlighting the differences to \cite{Boehi09a}.

Our experimental sequence starts by preparing a BEC in state $\ket{0}$ without a discernible thermal component in a static magnetic trap on the atom chip. Magnetic shielding and the use of stable current sources ensure stable magnetic potentials which allows us to prepare BECs with well-defined total atom number $N = 1250\pm 45$ through radio-frequency evaporative cooling.
The experiment is performed in a cigar-shaped magnetic trap at a distance of $44~\mu$m from the atom chip surface with longitudinal (axial) trapping frequency of $f_\mathrm{long}=109~$Hz ($f_\mathrm{ax}=500~$Hz) and a magnetic field in the trap center of $B_0=3.36~$G. 
%A simulation of the two-component Gross-Pitaevskii equation for our trap frequencies and atom number shows that spontaneous demixing of the two BEC components, as observed in \cite{Anderson09}, is negligible for our parameters.

\begin{figure}[tbh]
    \centering
        \includegraphics[width=1\columnwidth]{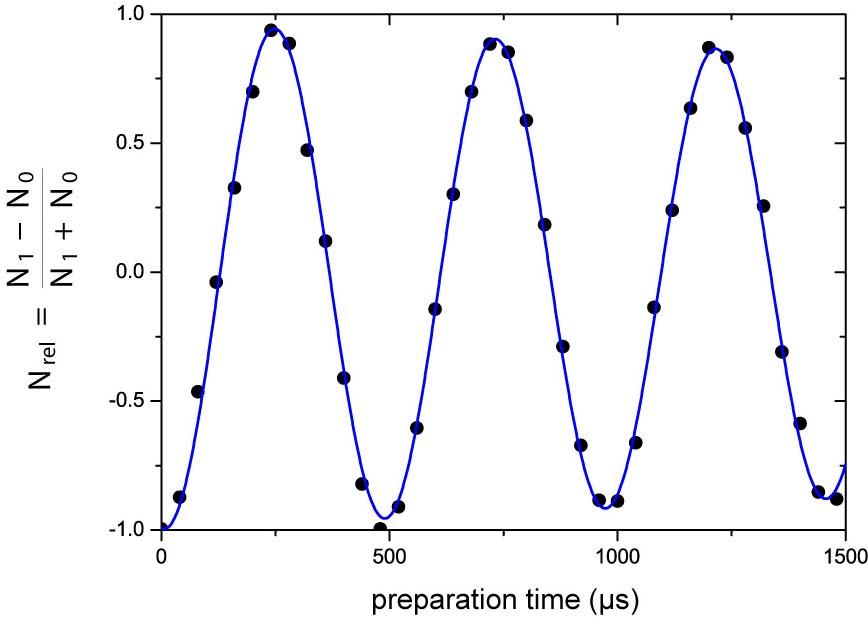}
    \caption{\textbf{Rabi oscillations.}
   Resonant Rabi oscillations of the relative atom number $N_\mathrm{rel}=(N_1-N_0)/(N_1+N_0)$ recorded  by varying the duration of the state preparation pulse. The efficiency of a $\pi$-pulse is $(96\pm 1)\,$\%. The decay with a time constant of $15~$ms is due to gradients in $\Omega$ near the structured metallic chip surface which imposes boundary conditions on the electromagnetic field.
     }
    \label{fig:Rabi}
\end{figure}

The BEC internal state is manipulated by coherently coupling the two-photon transition $\ket{0}\leftrightarrow \ket{1}$ with radio frequency and microwave radiation from an off-chip antenna and horn, respectively. The microwave is tuned $2\pi \times 360$~kHz above the transition to the intermediate state $\ket{F=2,m_F=0}$, resulting in a two-photon Rabi frequency of $\Omega=2\pi\times 2.1$~kHz. 
Fig.~\ref{fig:Rabi} shows the resulting Rabi oscillations for a detuning $\delta=0$ from two-photon resonance. The efficiency of a $\pi$-pulse is $(96\pm 1)\,$\%. % (or: the contrast is $(98\pm 1)\,$\%).

\begin{figure}[tbh]
    \centering
        \includegraphics[width=1\columnwidth]{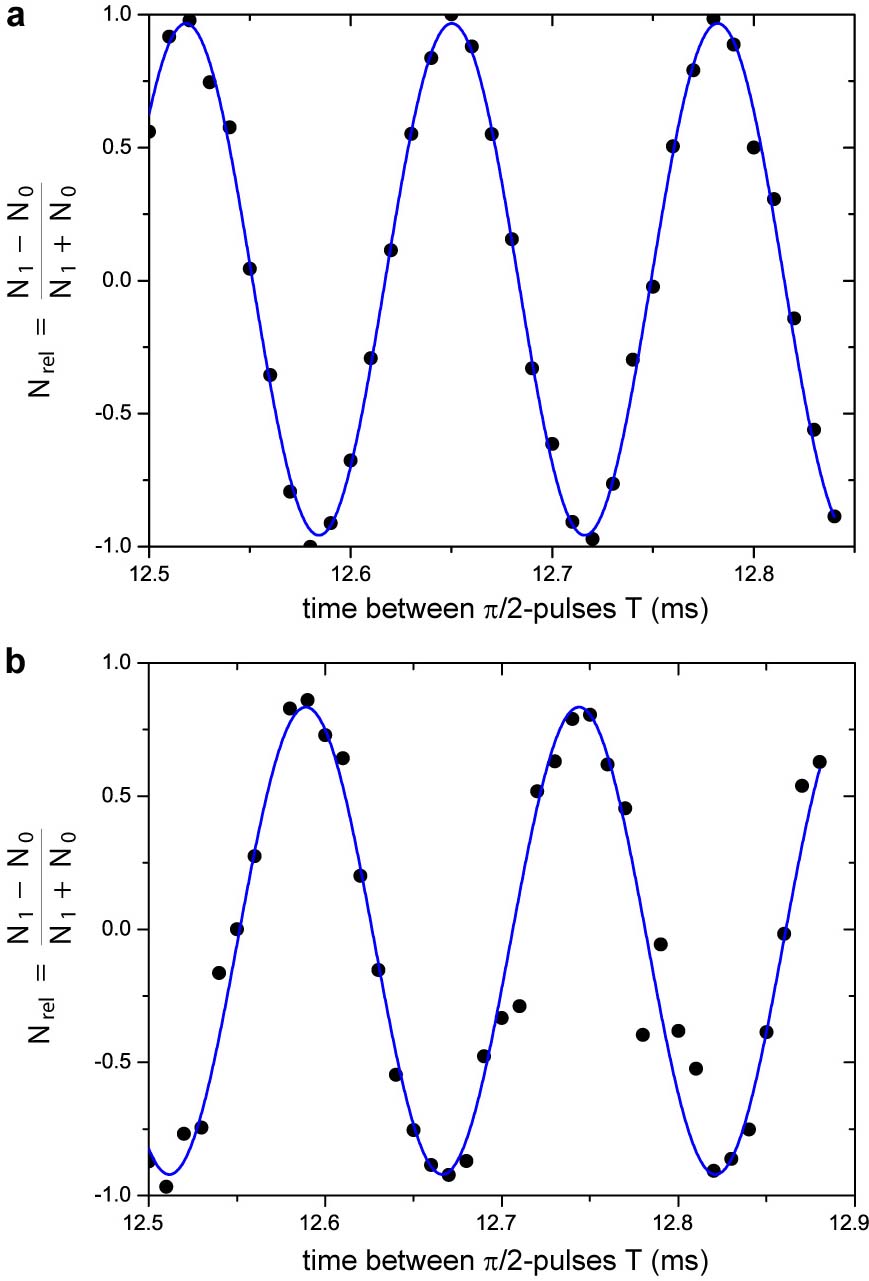}
    \caption{\textbf{Ramsey interference fringes.}
Ramsey interference fringes in the relative atom number $N_\mathrm{rel}$ recorded by varying the delay $T$ between two $\pi/2$-pulses. \textbf{a}, Ramsey fringes in the reference sequence in a static magnetic trap. The contrast obtained from a sinusoidal fit is $C=(96\pm 1)\,$\%. \textbf{b}, Ramsey fringes in the squeezing sequence with state-selective splitting and recombination of the BEC embedded between the $\pi /2$-pulses, as in Fig.~1d of the main text. The contrast is $C=(88\pm 3)\,$\%. 
     }
    \label{fig:Ramsey}
\end{figure}

Figure~\ref{fig:Ramsey}a shows Ramsey interference fringes between $\ket{0}$ and $\ket{1}$ as a function of the delay $T$ between two $\pi/2$-pulses. This is a reference measurement taken in a static magnetic trap without splitting the condensate, i.e.\ with the BEC in a coherent spin state.
The Ramsey contrast at $T=12.7~$ms is $C=(96\pm 1)\,$\%. %, demonstrating good control of the preparation pulses. 
%Extending such measurements to large $T$, we observe a contrast decay with a time constant of xxx ms, limited by 2-body losses in state $\ket{1}$ (???).
While the pulses are applied, the two-photon resonance frequency is shifted by $\Delta_\mathrm{mw}=2\pi\times 7.6\,$kHz with respect to the undriven system. This is due to differential AC Zeeman level shifts of $\ket{0}$ and $\ket{1}$ caused by the detuned microwave radiation of the two-photon drive \cite{Gentile89}.
We always adjust the frequency of the two-photon drive such that the detuning from two-photon resonance is $\delta=0$ while the pulse is applied. In between the pulses, the phase of the atomic superposition state thus evolves at a rate $-\Delta_\mathrm{mw}$ with respect to the two photon drive. This determines the frequency of the Ramsey oscillations in Fig.~\ref{fig:Ramsey}a.
%When the two states $\ket{0}$ and $\ket{1}$ are not coupled their relative phase therefore evolves at a rate $\omega$ and applying two $\pi /2$-pulses with a variable delay $T$ between them leads to Ramsey oscillations in the relative atom number at this rate.

For state-selective spatial splitting and recombination of the two BEC components, we use a microwave near-field potential created with an on-chip waveguide. Compared with \cite{Boehi09a}, the detuning of the microwave near-field from the transition $\ket{F=1,m_F=0}\leftrightarrow \ket{F=2,m_F=0}$ is much larger, $2\pi \times 12~$MHz, so that admixtures of other states to $\ket{0}$ and $\ket{1}$ are smaller and the states are more robust against magnetic field noise. In this configuration, both states experience a microwave potential of opposite sign, and with different magnitude due to the different hyperfine transition strengths. At a microwave power of $P_\mathrm{mw}=120$~mW launched into the chip, a splitting of the potential minima for the two states of $s = 0.52~\mu$m results. %, as determined by a simulation of the potential.

Figure~\ref{fig:Ramsey}b shows Ramsey fringes measured in the squeezing sequence, i.e.\ the BEC is split and recombined during the time $T$ between the $\pi/2$-pulses as in Fig.~1 of the main text.
Turning on the microwave near-field potential %during the time $T$ between the two $\pi /2$-pulses 
has two effects: The oscillation frequency of the Ramsey fringes slightly decreases because of the differential energy shift experienced by the two states in the potential. More importantly, the fringe contrast is modulated by the overlap of the BEC mode functions $\phi_0$ and $\phi_1$. 
The contrast at $T=12.7~$ms, the time at which the squeezed state is analyzed in the experiment, determines the length of the mean collective spin: $C= 2 |\langle S_x \rangle|/N$. We observe a contrast of $C=(88\pm 3)\,$\%, smaller than in the reference because the overlap of the BEC mode functions after splitting and recombination is less than unity. 
The difference to the contrast predicted by the dynamical simulation of $94\,$\% can most likely be explained by small motion in the transverse direction which is excited in the experiment but not modeled.
The contrast could be increased by optimal control of the atomic motion rather than abrupt switching of the potentials, as described in \cite{Treutlein06b}.

\subsection{Imaging system}

Compared with \cite{Boehi09a}, we use an improved detection system which allows us to take absorption images of both states and determine $N_0$ and $N_1$ in a single shot of the experiment. %This is used to calibrate out fluctuations in the total atom number $N=N_0+N_1$ (see Supplementary Information).
We use a back-illuminated deep-depletion CCD camera with a quantum efficiency of $90\,$\% at $780$~nm and fast line transfer.
After the squeezing sequence described in the main text, the atoms are transferred within $30~$ms into a relaxed trap with $f_\mathrm{long}=40~$Hz and $f_\mathrm{ax}=130~$Hz at a distance of $200~\mu$m from the chip surface.
The trap is switched off and states $\ket{1}$ and $\ket{0}$ are imaged after times-of-flight of $4.6~$ms and $6.1~$ms, respectively.
State $\ket{1}$ is directly imaged with a $\sigma^-$-polarized resonant laser beam on the $F=2\rightarrow F'=3$ cycling transition. After the image is taken, atoms in $\ket{1}$ fly out of the depth of focus of the imaging system due to the photon recoil momentum transferred during the imaging pulse. Subsequently, state $\ket{0}$ is optically pumped into the $F=2$ manifold of the ground state using a $F=1 \rightarrow F'=2$ pumping laser \cite{Matthews98} and imaged on the $F=2\rightarrow F'=3$ transition. 
For both states, the imaging pulse duration is $40~\mu$s and the imaging intensity is $I=0.8\,I_{\mathrm{sat}}$, where $I_\mathrm{sat}$ is the saturation intensity on the cycling transition.
The FWHM diameters of the imaged atom clouds are $15~\mu$m in the vertical and $10~\mu$m in the horizontal direction, both larger than the optical resolution of our imaging system of $4~\mu$m. The maximum optical densities in the cloud centers for 600 atoms in each state are 1.2 and 1.4, respectively.
The good agreement of the observed Rabi oscillations with the expected sinusoidal behavior 
proves %is an independent test of 
the linearity of our imaging system.

For the correct determination of the fluctuations of $S_\theta$ it is crucial to know the total atom number $N=N_0+N_1$ accurately.
We calibrate our imaging system by two independent methods. 
First, following the method of \cite{Reinaudi07}, we adjust the effective scattering cross section $\sigma_{\mathrm{eff}}$ which is used to calculate the atom number from the optical density of the cloud such that the measured atom number is independent of the imaging light intensity. 
We find $\sigma_{\mathrm{eff}}=0.9\,\sigma_0$, where $\sigma_0$ is the theoretically expected scattering cross section on the $\ket{F=2,m_F=-2}\leftrightarrow \ket{F'=3,m_F'=-3}$ cycling transition. This is plausible taking into account optical pumping in the beginning of the imaging pulse and imperfect imaging light polarization.

\begin{figure}[tbh]
    \centering
       \includegraphics[width=1\columnwidth]{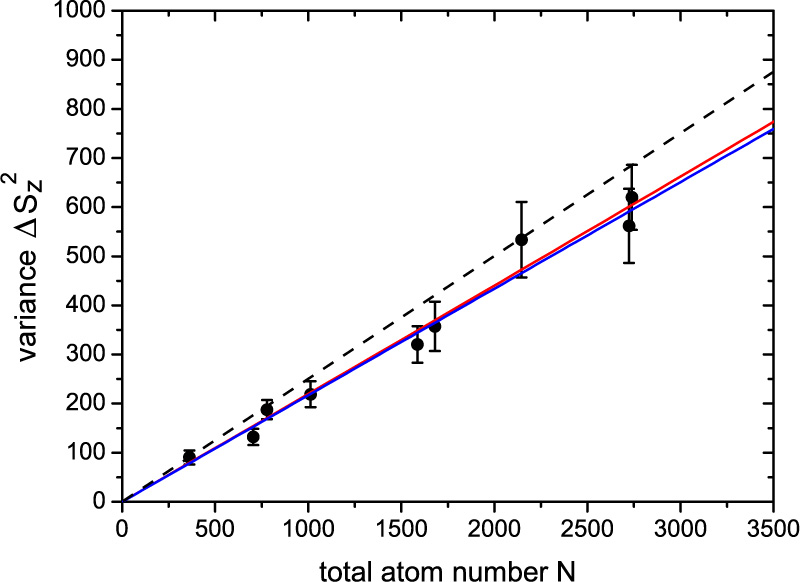}
    \caption{\textbf{Projection noise as a function of atom number.} A measurement of the variance $\Delta S_z^2$ directly after a $\pi/2$-pulse is shown as a function of total atom number $N$, where $N$ is independently calibrated with the method of \cite{Reinaudi07}. The dashed line shows the expected linear scaling with a slope of $1/4$, the blue line is a linear fit to the data which yields a slope of $0.22\pm 0.01$. The red line is a quadratic fit with $\Delta S_z^2 = a N + b N^2$. It yields $a=0.22\pm 0.02$ and $b=(0.06 \pm 1) \times 10^{-5}$, confirming the linear scaling expected for projection noise.
     }
    \label{fig:ProjNoise}
\end{figure}

An independent test of this calibration can be obtained by observing the scaling of projection noise with total atom number for a coherent spin state. 
Figure~\ref{fig:ProjNoise} shows the variance $\Delta S_z^2$ measured directly after a $\pi/2$-pulse as a function of the total atom number $N$. 
The constant offset due to imaging noise is subtracted as described below.
The observed linear behavior confirms that projection noise $\Delta S_z^2 \propto N$ dominates over technical noise which generically scales as $\Delta S_z^2 \propto N^2$.
If we use the first method to calibrate $N$, a fit to the data in Fig.~\ref{fig:ProjNoise} with a straight line through the origin yields a slope of $0.22\pm 0.01$. This agrees with the theoretically expected slope of $1/4$ to better than $15\,$\%. 
The difference lies within the error bar of our atom number calibration according to the first method. 
As the dependence of $\Delta S_z^2$ on $N$ can be very accurately determined from Fig.~\ref{fig:ProjNoise}, we use it to calibrate the total atom number by rescaling $N$ so that the slope of the linear fit is $1/4$.

\subsection{Data evaluation}
Our experiment runs very stable and we can usually let it record data overnight without supervision. The data shown in Fig.~2 of the main text is gathered during 7 measurement nights with between 80 (for $20 < \theta < 90$) and 370 (for $0 < \theta < 20$) experimental realizations per point. The error bars are statistical errors assuming a normal distribution.
%Nevertheless it can happen that the prepared atom number slightly drifts over night due to temperature drifts in the lab which influence laser locking and fiber coupling (other reasons? battery powered current sources? change of RF-stop-frequency because of drift in FM-input?) or that in a few shots no or only very few atoms are prepared. (What about drifts in the pulse area?)

We analyze the recorded images by counting atoms in the two states within two small rectangular regions with a typical size of $30~\mu\mathrm{m}\times 30~\mu\mathrm{m}$. This yields $N_0$ and $N_1$ and thus $S_z = (N_1-N_0)/2$. 
Only shots where the total atom number $N$ differs by no more than 150 from the mean $N=1250$ are used for the analysis and we check that a tighter (75 atoms) or wider (250 atoms) post-selection does not significantly change the data quality.
%To correct for different detection efficiency of states $\ket{0}$ and $\ket{1}$ we calculate $N_1^{\mathrm{det}}=N_1^{\mathrm{count}}*f$ and $N_0^{\mathrm{det}}=N_0^{\mathrm{count}}$ where $f$ is a factor between 1.05 and 1.15 which drifts slowly over time but is determined everyday anew by recording a Rabi-oscillation.
In the data for $90^\circ < \theta < 360^\circ$ a slow drift of $N_1-N_0$ is observed. We correct for this technical drift by subtracting a filtered data set % for $N_0$ and $N_1$ 
from the respective raw data, using a second order Savitzky-Golay filter \cite{Savitzky64} over 300 shots.

In Fig.~2a of the main text, we additionally correct $\Delta S_z^2$ for noise in our imaging system.
Photon shot noise from the imaging beam contributes to the measured fluctuations of $N_0$ and $N_1$ with a standard deviation of $\Delta N_{0,\mathrm{psn}}$ and $\Delta N_{1,\mathrm{psn}}$, respectively. We determine this noise either by measuring the apparent atom number fluctuations in regions of the image where no atoms are present, by taking `reference shots' where no atoms are prepared in the first place, or by calculating the expected shot noise from the observed imaging light intensity. All methods yield similar results.
We correct the variance of $S_z$ as
\[ \Delta S_{z,\mathrm{corr}}^2=\Delta S_z^2-(\Delta N_{0,\mathrm{psn}}^2+\Delta N_{1,\mathrm{psn}}^2)/4. \]
The applied correction due to imaging noise corresponds to typically $(\Delta N_{0,\mathrm{psn}}^2+\Delta N_{1,\mathrm{psn}}^2)^{1/2}/2 \approx 7$ atoms.
Finally, we normalize the variance to the expected variance for a coherent spin state with $\langle N_0\rangle=\langle N_1\rangle=\langle N\rangle/2$, obtaining $\Delta_n S_z^2=4\,\Delta S_{z,\mathrm{corr}}^2/\langle N\rangle$.
Without subtraction of imaging noise, we still observe a reduction in the spin fluctuations of $-2.3$~dB. %, which implies a metrologically useful squeezing of $-1.1$~dB.

\subsection*{Phase noise}
In our experiment, phase noise is the main reason why we do not reach the theoretically predicted spin noise reduction. It is therefore important to identify and eliminate the main technical sources for this noise.
In the squeezing sequence with $\theta=6^\circ$, where we observe the minimum of $\Delta S_\theta^2$, we measure the dependence of $S_\theta$ on various experimental parameters. 
From this we can calculate the sensitivity of the relative phase between the atomic state and our two-photon drive on these parameters.
%We measure the sensitivity of the relative phase of the atomic state and our two-photon drive at the time $T=12.7~$ms (in the squeezing sequence) to various experimental parameters. 
We independently determine the technical fluctuations of these parameters and estimate their contribution to the total phase noise of $\Delta \varphi =8^\circ$.
Timing jitter of our experiment control ($\approx 100~$ps), fluctuations of the external magnetic field (reduced by a $\mu$-metal shield to $\approx 0.3~$mG), power fluctuations ($\approx 5\times 10^{-3} $) and phase instabilities ($\approx 0.3^\circ$) of the radio frequency and microwave generators and amplifiers for the two-photon drive, and fluctuations of the on-chip currents together contribute a phase noise of about $1^\circ$. The microwave power coupled into the on-chip waveguide for creating the near-field potential fluctuates by $60~\mu$W ($5\times 10^{-4}$) leading to a phase noise of $2^\circ$.
The remaining phase noise is consistent with fluctuations of a current source used to create a magnetic field for the static magnetic trap. Its effect on the phase is not due to the small differential magnetic moment between $\ket{0}$ and $\ket{1}$ but due to a shift of the magnetic trap position in the highly inhomogeneous microwave near-field which leads to fluctuating microwave level shifts.

\subsection{Wigner Function reconstruction}
We reconstruct the Wigner function of the spin-squeezed condensate in the following way: For each measured $\theta \in [-90^\circ, 90^\circ]$ we create a histogram of $S_\theta$ and fit it with a cubic spline to obtain a smooth curve. We then use a filtered back-projection algorithm \cite{Natterer86} to perform an inverse Radon transform \cite{Radon17}.
The inverse Radon transform is derived for classical image reconstruction in a plane. 
In \cite{Breitenbach97}, it has been used to reconstruct the Wigner function of squeezed states of the electromagnetic field. However, it is generally not suited to reconstruct an arbitrary spin state on the curved Bloch sphere.
In our case the spin-squeezed state does not `wrap around' the Bloch sphere so that we can locally approximate the Bloch sphere by a plane and use the inverse Radon transform.
We furthermore make a continuum approximation to the measured values of $S_z$, which is reasonable as our imaging system does not have single atom resolution.
With the experimental method presented here but with a more sophisticated analysis \cite{DAriano03}, the density matrix of arbitrary spin states that spread over the whole Bloch sphere %(such as Schrödinger cat states)
can be reconstructed.
Quantum state tomography is of interest as it gives access to measures of entanglement 
%than $\xi^2$  which are not accessible by a simple measurements at one angle $\theta$,
such as the quantum Fisher information \cite{Frieden95, Pezze09}, which characterizes a more general class of states (including states with $\xi^2>1$) that can be used to overcome the standard quantum limit \cite{Pezze09}.

%
%\begin{figure}[tbh]
%    \centering
%        \includegraphics[width=1\columnwidth]{pulse_noise}
%    \caption{\textbf{Stability of the pulse area of the first pulse.}
%    Vary pulse length of first pulse in three steps: $\pi/2$, $3\pi/2$, and $5\pi/2$. Only a very small increase over shot noise is observed.
%     }
%    %\label{fig:PhaseNoise}
%\end{figure}
%

\end{document}